\documentclass[10pt,twocolumn,twoside]{IEEEtran}

\usepackage{amsmath}
\usepackage{amsthm}
\usepackage{epsfig}
\usepackage[noadjust]{cite}
\usepackage{amssymb}
\usepackage{psfrag}
\usepackage[latin1]{inputenc}
\usepackage{latexsym}
\usepackage[all]{xy}
\usepackage{subfigure}
\usepackage{color}
\usepackage{mathrsfs}
\newtheorem{theorem}{Theorem}
\newtheorem{lemma}{Lemma}

\begin{document}

\title{Multi-User Cooperative Diversity through Network Coding Based on Classical Coding Theory}

\author{João~Luiz~Rebelatto,~\IEEEmembership{Student~Member,~IEEE,} Bartolomeu~F.~Uchôa-Filho,~\IEEEmembership{Member,~IEEE,} Yonghui~Li,~\IEEEmembership{Senior Member,~IEEE,} and~Branka~Vucetic,~\IEEEmembership{Fellow,~IEEE}%
%\thanks{EDICS: WIN-CONT (Cooperative networking)}%
\thanks{Part of this work was presented at the IEEE International Symposium on Information Theory (ISIT'10), Austin, USA, June, 2010.}%
\thanks{João Luiz Rebelatto and Bartolomeu F. Uchôa-Filho are with the Communications Research Group, Department of Electrical Engineering, Federal University of Santa Catarina, Florianópolis, SC, 88040-900, Brazil (e-mail: \{jlrebelatto, uchoa\}@eel.ufsc.br).}%
\thanks{Yonghui Li and Branka Vucetic are with the School of Electrical \& Information Engineering, the University of Sydney, Sydney, NSW, 2006, Australia (e-mail: \{lyh, branka\}@ee.usyd.edu.au).}}%

%
%\author{\IEEEauthorblockN{João Luiz Rebelatto}
%\IEEEauthorblockA{Communications Research Group\\ Department of Electrical Engineering \\ Federal University of Santa Catarina\\ Florianópolis, SC, 88040-900, Brazil \\Email:  jlrebelatto@eel.ufsc.br}
%\and
%\IEEEauthorblockN{Bartolomeu F. Uchôa-Filho}
%\IEEEauthorblockA{Communications Research Group\\ Department of Electrical Engineering \\ Federal University of Santa Catarina\\ Florianópolis, SC, 88040-900, Brazil \\Email:  uchoa@eel.ufsc.br}
%\and
%\IEEEauthorblockN{Yonghui Li}
%\IEEEauthorblockA{Telecommunications Group\\ Department of Electrical Engineering \\ University of Sydney\\ NSW 2006, Australia\\Email:  lyh@ee.usyd.edu.au}}

%\ninept
%\topmargin=0mm

%\markboth{To be submitted to IEEE Transactions on Information Theory, Facets of Coding Theory: From Algorithms to Networks, 2010}%
%{Shell \MakeLowercase{\textit{et al.}}: Bare Demo of IEEEtran.cls for Journals}

\maketitle

\begin{abstract}
In this work, we propose and analyze a generalized construction of distributed network codes for a network consisting of $M$ users sending different information to a common base station through independent block fading channels. The aim is to increase the diversity order of the system without reducing its throughput. The proposed scheme, called generalized dynamic-network codes (GDNC), is a generalization of the dynamic-network codes (DNC) recently proposed by Xiao and Skoglund. The design of the network codes that maximize the diversity order is recognized as equivalent to the design of linear block codes over a nonbinary finite field under the Hamming metric. We prove that adopting a systematic generator matrix of a maximum distance separable block code over a sufficiently large finite field as the network transfer matrix is a sufficient condition for full diversity order under link failure model. The proposed generalization offers a much better tradeoff between rate and diversity order compared to the DNC. An outage probability analysis showing the improved performance is carried out, and computer simulations results are shown to agree with the analytical results.
\end{abstract}
\IEEEpeerreviewmaketitle
\section{Introduction} \label{sec:introduction}

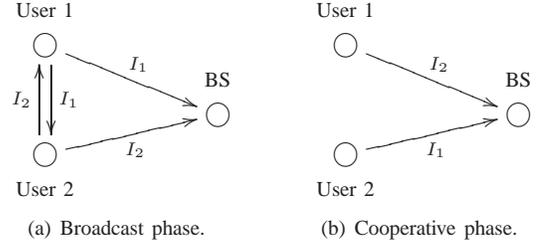
\begin{figure}[!t]
\center{
\hfill
\subfigure[Broadcast phase.\label{fig:DAF_BC}]{
\xymatrix@C=1.5cm@-2pc{
\text{\footnotesize User 1} & & \\
\bigcirc \ar@{->}[rrdd]^-{I_1} \ar@{->}@<0.5ex>[ddd]^-{I_1}& &                 \\
& & \text{{\footnotesize BS}}\\
& &  \bigcirc\\
\bigcirc \ar@{->}[rru]_-{I_2}  \ar@{->}@<0.5ex>[uuu]^-{I_2}  & & \\
\text{\footnotesize User 2} & &   \\  \\
}}\hfill
\subfigure[Cooperative phase.\label{fig:DAF_COOP}]{
\xymatrix@C=1.5cm@-2pc{
\text{\footnotesize User 1} & & \\
\bigcirc \ar@{->}[rrdd]^-{I_2}  & &                 \\
& & \text{{\footnotesize BS}}\\
& &  \bigcirc\\
\bigcirc \ar@{->}[rru]_-{I_1}  & & \\
\text{\footnotesize User 2} & &          \\        \\
}}\hfill}
\caption{Two-user decode-and-forward cooperative network. (a) Each user broadcasts its own information and (b) each user transmits its partner's information after decoding and re-encoding it.}
\label{fig:DAF}
\end{figure}

\IEEEPARstart{I}{n} a cooperative wireless communications system with multiple users transmitting independent information to a common base station (BS), besides broadcasting their own information, users help each other relaying their partners information~\cite{sendonaris.03,laneman.04,hunter.02,xiaoL.07,xiao.09,xiao.10,bing.08}. In the \emph{decode-and-forward} (DAF) relaying protocol~\cite{cover.79}, the codeword relayed to the base station (BS) is a re-encoded version of the previously decoded codeword received in the broadcasting phase. A simple 2-user DAF network is illustrated in Fig.~\ref{fig:DAF}. In the first time slot (Fig.~\ref{fig:DAF_BC}), each user broadcasts its own information (usually through orthogonal channels). In the second time slot (Fig.~\ref{fig:DAF_COOP}), each user transmits its partner's information to the BS after decoding and re-encoding it. If the user cannot decode the partner's information, it sends its own information again. By transmitting the same information through independent channels, a cooperative diversity is achieved~\cite{sendonaris.03,laneman.04}.

Network coding~\cite{ahlswede.00,koeter.03}, a method originally proposed to attain maximum information flow in a network, has recently been applied to cooperative wireless communications systems to improve the bit error rate (BER) performance~\cite{xiaoL.07,xiao.09.ISIT,xiao.09,xiao.10,hausl.06,hausl.PhD.08,bing.08,zhan.08}. In a network coded system, relays process information from different users and perform linear combinations of the received signals, with coefficients chosen from a finite field GF$(q)$.

A two-user cooperative system that employs binary network coding (BNC) was proposed in~\cite{xiaoL.07}. In that scheme, each user transmits the binary sum (XOR) of its own source message and the message received from its partner (if correctly decoded). However, the scheme proposed in~\cite{xiaoL.07} does not improve the system diversity order.

In~\cite{xiao.09.ISIT}, it was shown that binary network coding is not optimal for achieving full diversity in multiple user-multiple relay systems when link failures are taken into account. A similar result was shown in~\cite{xiao.09}, however, instead of considering dedicated relays, therein the $M$ users themselves act as relays for each other. The scheme proposed in~\cite{xiao.09}, called \emph{dynamic-network codes} (DNC), considers a fixed nonbinary network code. In the first time slot, each user broadcasts a single packet of its own to the BS as well as to the other users, which try to decode the packet. From the second time slot until the $M$th time slot, each user transmits to the BS $M-1$ nonbinary linear combinations of the packets that it could successfully decode. With DNC, by using an appropriately designed network code,  the diversity order was shown to be higher than that in binary network coded systems. The scheme is called ``dynamic'' in the sense that the network code is designed to perform well under a possible occurrence of errors in the channels between the users (inter-user channels). However, this design approach can become extremely complex as the number of users $M$ grows, as shown in~\cite{xiao.10}.

In~\cite{yeung.06a,yeung.06b}, Yeung and Cai derived a generalization of the Hamming, the Singleton and the Gilbert-Varshamov bounds, extending concepts from the classical point-to-point coding theory to a communication network, and focusing on its error correction capability.  In~\cite{zhang.08}, it was shown that the minimum distance of a network error-correction code for a single-source multicast case plays the same role as it plays in classical coding theory. The decoding performance under the occurrence of different types of errors and erasures was also evaluated.

 In this paper, we investigate another relationship between network codes and classical error-correcting codes. We elaborate on the DNC scheme by first recognizing the problem as equivalent to that of designing linear block codes over GF($q$) for erasure correction. In particular, for perfect inter-user channels, we note that the diversity order equals the minimum Hamming distance of the block code, so the network transfer matrix should correspond to the generator matrix of an optimal block code under the Hamming metric. The Singleton bound appears as a natural upper bound on the diversity order, and this bound is achieved with a sufficiently large field size~\cite{sloane.77}.

We then extend the DNC scheme by allowing each user to broadcast several (as opposed to just one) packets of its own in the broadcast phase, as well as to transmit several nonbinary linear combinations (of all correctly decoded packets) in the cooperative phase. From the block coding perspective, the so-called {\em generalized dynamic-network codes} (GDNC) consider a longer codeword, with more parity symbols, which improves the Singleton bound.

Codes that achieve the Singleton bound are called {\it maximum distance separable} (MDS) codes~\cite{sloane.77}. Regarding the GDNC network code design, we show that if a generator matrix of a MDS code is used as the GDNC network code, the maximum diversity order is guaranteed. We also show that a much better tradeoff between rate and diversity order can be achieved, {\em e.g.}, it is possible to achieve both rate and diversity order higher than those in the DNC scheme.

%Finally, we evaluate the performance of the GDNC scheme making use of the diversity-multiplexing tradeoff (DMT) introduced by Zheng and Tse in~\cite{zheng.03} for point-to-point multiple-input multiple-output (MIMO) channels and later applied to the (noncooperative) multiple-access channel in~\cite{tse.04}. We show that our scheme outperforms other schemes previously proposed (such as the Non-orthogonal amplify-and-forward (NAF) and Cooperative Multiple Access (CMA) protocols presented in~\cite{azarian.05}) in a large range of code rates, from moderate to high.

The rest of this paper is organized as follows. The next section presents the system model and some relevant previous works, including binary network coded cooperative systems~\cite{xiaoL.07} and the DNC scheme~\cite{xiao.09}. The motivation for GDNC is presented in Section~\ref{sec:motivation}. Section~\ref{sec:gdnc} presents the proposed GDNC scheme, showing firstly a simple 2-user network and then generalizing it to $M$ users. An outage probability analysis and our network code design approach are carried out in Sections~\ref{sec:outage} and~\ref{sec:net_design}, respectively. Simulations results are presented in Section~\ref{sec:simulations}. Finally, Section~\ref{sec:conclusions} presents our conclusions and final comments.

\section{Preliminaries} \label{sec:preliminaries}

\subsection{System Model}

The system consists of a multiple access channel (MAC) network, in which multiple users ($M \geq 2$) have different information to send to a common base station (BS).
One time slot (TS) is defined as the time period in which all the $M$ users perform a single transmission (through orthogonal channels, either in time, frequency or code). That is, one TS corresponds to $M$ transmissions from individual users.
The received baseband codeword at user $i$ at time $t$ is given by
\begin{equation}
y_{j,i,t} = h_{j,i,t}x_{j,t} + n_{j,i,t},
\end{equation}
where $j \in \{1,\cdots,M\}$ represents the transmit user index and $i \in \{0,1,\cdots,M\}$ the receive user index (0 corresponds to the BS). The index $t$ denotes the time slot. $x_{j,t}$ and $y_{j,i,t}$ are the transmitted and the received codewords, respectively. $n_{j,i,t}$ is the zero-mean additive white Gaussian noise with variance $N_0/2$ per dimension. The channel gain due to multipath fading is denoted by $h_{j,i,t}$, and it is assumed to have independent identically distributed (i.i.d.) (across space and time) Rayleigh distribution with unit variance.

Assuming $x_{j,t}$'s to be i.i.d. Gaussian random variables and considering all the channels with the same average  signal-to-noise ratio (SNR), the mutual information $I_{j,i,t}$ between $x_{j,t}$ and $y_{j,i,t}$ is
\begin{equation} \label{eq:mutual_inf}
I_{j,i,t} = \log(1+|h_{j,i,t}|^2\text{SNR}).
\end{equation}

In this work, we do not consider a joint network-channel code design \cite{hausl.06,hausl.PhD.08,bing.08}. It is assumed that powerful enough channel codes are being used at a link level, so that $x_{j,t}$ can be correctly decoded at user $i$ if $I_{j,i,t}>r_{j,i,t}$, where $r_{j,i,t}$ is the information rate from User $j$ to User $i$ in the time slot $t$. Considering that all the users have the same rate, the index of $r$ can be dropped. Thus, $x_{j,t}$ cannot be correctly decoded at user $i$ if
\begin{equation}
|h_{j,i,t}|^2<g,
\end{equation}
where $g=\frac{2^r-1}{\text{SNR}}$. The probability that such an event happens is called the \emph{outage probability}. For Rayleigh fading, the outage probability is calculated as~\cite{laneman.04,tse.05}
\begin{equation}
P_e = \Pr\left\{|h_{j,i,t}|^2<g \right\}=1-e^{-g}\approx g,
\end{equation}
where the approximation holds for a high SNR region. Considering block fading, the diversity order $D$ is defined as~\cite{tse.05}
\begin{equation} \label{eq:diversity}
D \triangleq \lim_{\text{SNR}\rightarrow \infty} \frac{-\log P_o}{\log \text{SNR}},
\end{equation}
where $P_o$ is the system overall outage probability. %It has been shown \cite{laneman.04} that for Fig. \ref{fig:DAF} the maximum achievable diversity order is $D=2$, under the assumption of quasi-static channels.

In this work, block fading means that fading coefficients are i.i.d. random variables for different blocks but constant during the same block. It is also assumed that the receivers have perfect channel state information (CSI), but the transmitters do not have any CSI.

Before we continue, we establish that throughout this paper, the operators $+$ and $-$ represent operations over real numbers, $\oplus$ is the binary addition (XOR), and  $\boxplus$ and $\boxminus$ are the addition and subtraction operations over a nonbinary field, respectively.

\subsection{Binary Network Coded Cooperation} \label{subsec:2user}

Instead of transmitting only the partner's information in the second TS, as shown in Fig.~\ref{fig:DAF}, each user of this 2-user system can transmit a binary sum of its own information and its partner's information. Considering for the moment that the two inter-user channels are reciprocal ({\em i.e.}, $h_{j,i,t}=h_{i,j,t}$), if the inter-user channel is not in outage, which happens with probability $(1-P_e)$, the four different packets transmitted to the BS are $I_1$, $I_2$, $I_1\oplus I_2$ and $I_1\oplus I_2$. Upon receiving two copies of the same message, the BS performs maximum ratio combining (MRC). In this case, an outage for $I_1\oplus I_2$ occurs if $(|h_{1,0,2}|^2+|h_{2,0,2}|^2)<g$ and has probability $P_1\approx \frac{P_e^2}{2}$ \cite{laneman.04}.

Without loss of generality, we analyze the outage probability of User 1. The same result holds for User 2 due to symmetry. An outage event occurs when the direct transmission of $I_1$ and one or two packets from $I_2$ and $I_1\oplus I_2$ cannot be decoded by the BS. In this case, the outage probability for User 1 is
\begin{equation}
P_{p,1} \approx P_e(P_e+P_1). \nonumber
\end{equation}
If the inter-user channel is in outage, which happens with probability $P_e$, each user just retransmits its own information. Upon receiving two copies of the same information packet, the BS performs MRC, resulting in the following outage probability for User 1~\cite{laneman.04}\[P_{f,1} \approx \frac{P_e^2}{2}.\] Considering all the possibilities for the inter-user channel, the outage probability of User 1 is then given by \[P_{o,1}=(1-P_e)P_{p,1}+P_eP_{f,1} \approx P_e^2.\]

If the inter-users channels are non-reciprocal, it can be shown that the outage probability of a 2-user binary network coded system is given by~\cite{xiaoL.07}
\begin{equation} \label{eq:out_net_bin}
P_{o,\text{BNC}}\approx P_e^2,
\end{equation}
which corresponds to a diversity order $D=2$ according to \eqref{eq:diversity}. We can see that the diversity order obtained from \eqref{eq:out_net_bin} is not optimal, since the information of each user is transmitted through three independent paths and higher diversity order can be achieved, as will be explained later. By a similar analysis, the outage probability of the 2-user DAF scheme, presented in Fig.~\ref{fig:DAF}, considering non-reciprocal inter-user channels, can be shown to be
\begin{equation}
P_{o,\text{DAF}} \approx 1.5P_e^2.
\end{equation}

%Instead of transmitting only the partner's information in the second TS, as in the DAF scheme, each user can transmit a binary sum of its own information and its partner's information. It was shown in~\cite{xiaoL.07} that the outage probability for a 2-user binary network coded system is given by
%\begin{equation} \label{eq:out_net_bin}
%P_{o,\text{BNC}}\approx P_e^2,
%\end{equation}
%which corresponds to a diversity order $D=2$ according to \eqref{eq:diversity}. We can see that the diversity order obtained from \eqref{eq:out_net_bin} is not an optimal result, since the information of each user is transmitted through three independent paths and higher diversity order can be achieved, as will be explained later. By a similar analysis, the outage probability of the DAF scheme is given by~\cite{laneman.04}
%\begin{equation}
%P_{o,{\text{DAF}}} \approx 0.5P_e^2.
%\end{equation}

\subsection{Dynamic-Network Codes} \label{subsec:xiao}

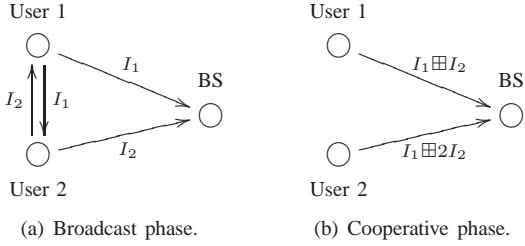
\begin{figure}[!t]
\center{
\hfill
\subfigure[Broadcast phase.\label{fig:xiaoa}]{
\xymatrix@C=1.5cm@-2pc{
\text{\footnotesize User 1} & & \\
\bigcirc \ar@{->}[rrdd]^-{I_1} \ar@{->}@<0.5ex>[ddd]^-{I_1}& &                 \\
& & \text{{\footnotesize BS}}\\
& &  \bigcirc\\
\bigcirc \ar@{->}[rru]_-{I_2}  \ar@{->}@<0.5ex>[uuu]^-{I_2}  & & \\
\text{\footnotesize User 2} & &   \\  \\
}}\hfill
\subfigure[Cooperative phase.\label{fig:xiaob}]{
\xymatrix@C=1.5cm@-2pc{
\text{\footnotesize User 1} & & \\
\bigcirc \ar@{->}[rrdd]^-{I_1\boxplus I_2}  & &                 \\
& & \text{{\footnotesize BS}}\\
& &  \bigcirc\\
\bigcirc \ar@{->}[rru]_-{I_1 \boxplus 2I_2}  & & \\
\text{\footnotesize User 2} & &          \\        \\
}}\hfill}
\caption{Two-user cooperative network with nonbinary network coding. (a) Each user broadcasts its own information and (b) each user transmits a linear combination over GF(4) of all the available information packets.}
\label{fig:Xiao}
\end{figure}

In \cite{xiao.09,xiao.10}, Xiao and Skoglund showed that the use of nonbinary network coding is necessary to achieve a higher diversity order, and then proposed the so-called DNC.
A simple 2-user DNC scheme is presented in Fig.~\ref{fig:Xiao}, where nonbinary coefficients are used.

Assuming reciprocal inter-user channels and considering that they are not in outage, which happens with probability $(1-P_e)$, the BS receives $I_1$, $I_2$, $I_1\boxplus I_2$ and $I_1\boxplus 2I_2$. We can see that the BS is able to recover the original messages $I_1$ and $I_2$ from any 2 out of the 4 received packets. For User 1 (the same result holds for User 2 due to symmetry), an outage occurs when the direct packet $I_1$ and at least 2 out of the 3 remaining packets cannot be decoded. This occurs with probability~\cite{xiao.09}
\begin{equation}
P_{p,1}=P_e\left[{{3}\choose{2}}P_e^2(1-P_e)+P_e^3\right] \approx 3P_e^3. \nonumber
\end{equation}

However, when considering the link failure of inter-user channels, with probability $P_e$ a user cannot decode its partner's message. In this situation, it retransmits its own information. The BS performs MRC, and the outage probability for User 1 is then  $P_{f,1} = \frac{P_e^2}{2}$~\cite{laneman.04}. Thus, the overall outage probability for User 1 is~\cite{xiao.09}
\begin{equation}
P_{o,1}= P_eP_{f,1}+(1-P_e)P_{p,1} \approx 3.5P_e^3. \nonumber
\end{equation}

It is easy to see that the diversity order is $D=3$. If the inter-user channels are non-reciprocal, then the outage probability  was shown to be~\cite{xiao.09}
\begin{equation} \label{eq:DNC_not_rec_M2}
P_{o,1}\approx 4P_e^3. \nonumber
\end{equation}

For $M$ users, in the DNC scheme, each user transmits a fixed number of $M-1$ nonbinary linear combinations in the cooperative phase.  It is shown in~\cite{xiao.09} that the diversity order achieved by DNC is
\begin{equation} \label{eq:dnc_diversity}
D_{\text{DNC}}=2M-1,
\end{equation}
however, with a low, fixed rate $R=M/M^2=1/M$.

The network code necessary to achieve the diversity order in~\eqref{eq:dnc_diversity} was proved to exist through an association between the DNC scheme and the single-source linear robust network codes \cite[Th.11]{koeter.03}. The network code design proposed in~\cite{xiao.09} is to find the transfer matrix which is nonsingular for all possible  error patterns in the inter-user channels. This network design method can have high complexity as $M$ increases.

In~\cite{xiao.10}, a simplified construction of DNCs was presented. This reduction in the design complexity is obtained at the expense of a poorer performance, which is negligible if the field size is as large as~\cite{xiao.10}
\begin{equation} \label{eq:q_dnc}
q \geq {{M^2-1}\choose{M-1}}.
\end{equation}

\section{Motivation for this Work} \label{sec:motivation}

The multiple-source-one-destination network can be represented by a transfer matrix which in turn can be seen as a generator matrix of a systematic linear block code. The generator matrix obtained from the 2-user system, illustrated in Fig.~\ref{fig:Xiao}, is given by
\begin{equation} \label{eq:G_DNC}
\textbf{G}_{\text{DNC}} =
\left[
\begin{array}{cc|cc}
1 & 0 & 1 & 1 \\
0 & 1 & 1 & 2 \\
\end{array} \right].
\end{equation}

In the DNC scheme~\cite{xiao.09,xiao.10}, the diversity order is related to the minimum number of correctly received packets (or symbols) at the BS from which the information packets from all users can be recovered. A package which is not received correctly may be thought of as an erasure, and is discarded by the receiver. The receiver's ability to recover the information packets from the non-erased packages is thus equivalent to the erasure correction capability of the associated (network) block code. It is well-known that the transmitted codeword of a linear block code with minimum Hamming distance $D$ can be recovered if no more than $D-1$ of its positions have been erased by the channel~\cite{costello.04}. The connection between these two problems establishes that the diversity order of the 2-user DNC system shown in Fig. \ref{fig:Xiao}, under the assumption of perfect inter-user channels, is equal to the minimum Hamming distance of the rate 2/4 block code with generator matrix given in \eqref{eq:G_DNC}.

In general, for a rate $k/n$ linear block code, the minimum Hamming distance is upper bounded by the Singleton bound~\cite{costello.04}:
 \begin{equation} \label{eq:singleton}
 d_{\text{min}} \leq n-k+1.
 \end{equation}
  As shown in~\cite[Chap.11, Coroll.7]{sloane.77}, the Singleton bound can only be achieved if the alphabet size is large enough. For example, for a $4/8$ block code, the Singleton bound gives $d_{\text{min}}\leq 5$. However, the maximum possible achieved minimum Hamming distance in GF$(2)$ is 3. In GF$(4)$, it is possible to achieve $d_{\text{min}}=4$. The upper bound $d_{\text{min}}=5$ is only possible to be achieved if the field size is at least 8~\cite{grassl.10}.

In the DNC scheme~\cite{xiao.09,xiao.10}, the overall rate is $M/M^2$. From \eqref{eq:singleton}, the diversity order is thus upper bounded by $D_{\max,\text{DNC}}=M^2-M+1$. We observe that a system with rate $R=\alpha M/ \alpha M^2$ (for $\alpha \geq 2$) would have the same overall rate as the DNC scheme ($R=1/M$), but the diversity upper bound would be increased to
\begin{equation} \label{eq:singleton_new}
D_{\max,\alpha}=\alpha (M^2-M)+1.
\end{equation}
This motivates us to modify the DNC scheme accordingly and to propose an even more general scheme which is more flexible in terms of rate and diversity order.

Due to inter-user channel outages, the upper bound presented in~\eqref{eq:singleton_new} cannot be achieved. In Section \ref{sec:outage}, the discrepancy between the upper bound and the real diversity order obtained is quantified by the outage probability analysis.

\section{Generalized Dynamic-Network Codes} \label{sec:gdnc}

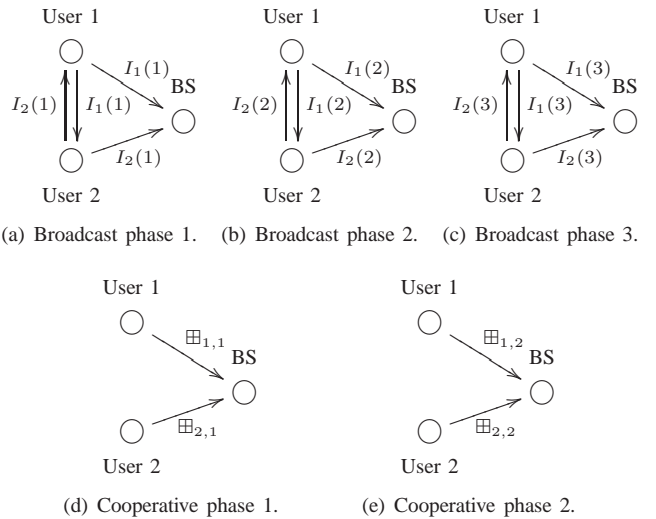
\begin{figure}[!t]
\centering
\subfigure[Broadcast phase 1.\label{fig:newa}]{
\xymatrix@C=1.1cm@-2pc{
\text{\footnotesize User 1} & & \\
\bigcirc \ar@{->}[rrdd]^-{I_1(1)} \ar@{->}@<0.5ex>[ddd]^-{I_1(1)}& &                 \\
& & \text{{\footnotesize BS}}\\
& &  \bigcirc\\
\bigcirc \ar@{->}[rru]_-{I_2(1)}  \ar@{->}@<0.5ex>[uuu]^-{I_2(1)}  & & \\
\text{\footnotesize User 2} & &   \\  \\
}}
\subfigure[Broadcast phase 2.\label{fig:newb}]{
\xymatrix@C=1.1cm@-2pc{
\text{\footnotesize User 1} & & \\
\bigcirc \ar@{->}[rrdd]^-{I_1(2)} \ar@{->}@<0.5ex>[ddd]^-{I_1(2)}& &                 \\
& & \text{{\footnotesize BS}}\\
& &  \bigcirc\\
\bigcirc \ar@{->}[rru]_-{I_2(2)}  \ar@{->}@<0.5ex>[uuu]^-{I_2(2)}  & & \\
\text{\footnotesize User 2} & &   \\  \\
}}
\subfigure[Broadcast phase 3.\label{fig:newc}]{
\xymatrix@C=1.1cm@-2pc{
\text{\footnotesize User 1} & & \\
\bigcirc \ar@{->}[rrdd]^-{I_1(3)} \ar@{->}@<0.5ex>[ddd]^-{I_1(3)}& &                 \\
& & \text{{\footnotesize BS}}\\
& &  \bigcirc\\
\bigcirc \ar@{->}[rru]_-{I_2(3)}  \ar@{->}@<0.5ex>[uuu]^-{I_2(3)}  & & \\
\text{\footnotesize User 2} & &   \\  \\
}} \\
\subfigure[Cooperative phase 1.\label{fig:newd}]{
\hspace{0.5cm}
\xymatrix@C=1.1cm@-2pc{
\text{\footnotesize User 1} & & \\
\bigcirc \ar@{->}[rrdd]^-{\boxplus_{1,1}}  & &                 \\
& & \text{{\footnotesize BS}}\\
& &  \bigcirc\\
\bigcirc \ar@{->}[rru]_-{\boxplus_{2,1}}  & & \\
\text{\footnotesize User 2} & &          \\        \\
}
\hspace{0.5cm}}
\hspace{0.2cm}
\subfigure[Cooperative phase 2.\label{fig:newe}]{
\hspace{0.5cm}
\xymatrix@C=1.1cm@-2pc{
\text{\footnotesize User 1} & & \\
\bigcirc \ar@{->}[rrdd]^-{\boxplus_{1,2}}  & &                 \\
& & \text{{\footnotesize BS}}\\
& &  \bigcirc\\
\bigcirc \ar@{->}[rru]_-{\boxplus_{2,2}}  & & \\
\text{\footnotesize User 2} & &          \\        \\
}
\hspace{0.5cm}}
\caption{A rate 6/10 GDNC scheme with $M=2$ users. The symbol $\boxplus_{m,t'}$ denotes the linear combination of all the available information packets performed by relay $m$ in time slot $t'$ of the cooperative phase.} %For a 6/10 block code, according to \cite{grassl.10}, the minimum field size necessary to achieve $d_{min}=5$ is $q=16$.}
\label{fig:New}
\end{figure}

We begin the description of the proposed GDNC scheme by elaborating on the 2-user DNC scheme presented in Fig.~\ref{fig:Xiao}, whose associated generator matrix is given in \eqref{eq:G_DNC}. The new scheme is illustrated in Fig.~\ref{fig:New}.

Each user broadcasts three packets (or symbols) of its own in the broadcast phase, and then each user transmits two nonbinary linear combinations (of the six previously broadcasted packets) over GF($q=2^b$) in the cooperative phase, where $b$ is an integer greater than zero.  The receiver collects the 10 packets, which can be seen as a codeword of a systematic 6/10 linear block code.

Without loss of generality, we analyze the outage probability for User 1 in the first TS. Let us assume for the moment that the inter-user channels are reciprocal. In this case, the probability that no errors occur in the inter-user channel is $1-P_e$. If User 2 can correctly decode $I_1(1)$, the message of User 1 in the time slot 1, then $I_1(1)$ will be in outage at the BS if the direct packet \emph{and}, in the worst case, the 4 parity packets containing $I_1(1)$ cannot be decoded by the BS\footnote{If User 2 can correctly decode $I_1(1)$, there are other situations yielding the same outage probability for message $I_1(1)$. But, for the moment being, we are not interested in the multiplicity of the outage patterns, but only in the diversity order.}, which happens with probability $P_{p,1} \approx P_e^5$.

%Without loss of generality, we analyze the outage probability for User 1 in the first TS. Consider for the moment being that the inter-user channels are reciprocal. In this case, the probability that no errors occur in the inter-user channel is $P_{p,1}=1-P_e$. If User 2 can correctly decode\footnote{Here, we do not consider errors in the decoding process. If a valid codeword is obtained after the decoding, it is assumed to be the correct codeword.} $I_1(1)$, the message $I_1(1)$ will suffer an outage at the BS when the direct packet \emph{and}, in the worst case, the 4 parity packets containing $I_1(1)$ cannot be decoded by the BS, which happens with probability $P_0 \approx P_e^5$.

If User 2 cannot correctly decode $I_1(1)$, which happens with probability $P_e$, it will not be able to help User 1 by relaying $I_1(1)$. In this case, the BS receives only 3 packets containing $I_1(1)$ (the direct transmission plus 2 parity-checks sent by User 1 itself). Message $I_1(1)$ will then be in outage at the BS if the direct packet \emph{and}, in the worst case, both of the parity-checks transmitted by User 1 cannot be decoded either. This happens with probability $P_{f,1} \approx P_e^3$. Therefore, considering reciprocal inter-user channels and all the outage patterns, the outage probability of message $I_1(1)$ is given by
\begin{equation} \label{eq:new_2user_outage}
P_{o,1} = P_e P_{f,1}+(1-P_e)P_{p,1} \approx P_e^4.
\end{equation}
We can show that the same result is obtained when the inter-user channels are non-reciprocal.

We can see from~\eqref{eq:new_2user_outage} that the diversity order achieved by the rate 6/10 GDNC scheme with $M=2$ users presented in Fig.~\ref{fig:New} is $D=4$, which is higher than the one obtained by the rate 2/4 DNC scheme of~\cite{xiao.09,xiao.10} with $M=2$ users in \eqref{eq:DNC_not_rec_M2}. Therefore, both the rate and the diversity order have increased.

We can also verify that the outage probability is dominated by the term related to the inter-user channel being in outage, when User 2 cannot help User 1. Similarly, for a $M$-user network, we will see that when all the $M-1$ inter-user channels are in outage is also the worst-case scenario for the outage probability of a given message.

\subsection{Multiple Users} \label{subsec:new_mult}

\begin{figure}[!t]
\[
\xymatrix@-2pc{
\txt{User $1$:}& & I_1(1)& \cdots  & I_1(k_1)&\ar@{--}[dddd] & P_1(1)& \cdots  & P_1(k_2)  \\
\txt{User $2$:}& & I_2(1)& \cdots  & I_2(k_1)& & P_2(1)& \cdots  & P_2(k_2) \\
& &\vdots & \ddots  & \vdots & &\vdots & \ddots  & \vdots \\
\txt{User $M$:}& & I_M(1)& \cdots & I_M(k_1)& & P_M(1)& \cdots & P_M(k_2)  \\
 & \ar[rrrrrrrr]_-{\text{time slots}} & & &&&&&&\\
}
\]
\caption{GDNC scheme for a network with $M$ users.}
\label{fig:Gen_proposed}
\end{figure}
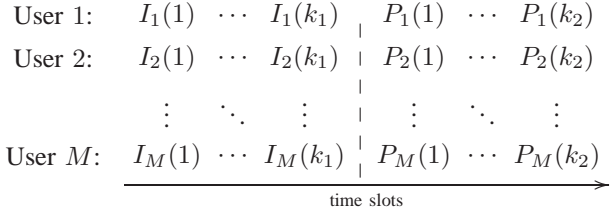

The generalization of the scheme presented in Fig.~\ref{fig:New} to the case of $M > 2$ users is illustrated in Fig.~\ref{fig:Gen_proposed}. $I_j(t)$ represents the information (symbol or packet) transmitted  by user $j$ ($j=1,\cdots,M$) in time slot $t$ ($t=1,\cdots,k_1$) of the broadcast phase (at left), and $P_j(t')$ corresponds to the parity-check transmitted (symbol or packet) by user $j$ in time slot $t'$ ($t'=1,\cdots,k_2$) of the cooperative phase (at right).

Following Fig.~\ref{fig:Gen_proposed}, each user first broadcasts $k_1$ independent information packets of its own\footnote{These packets can be sent in any order. Recall that the channels are considered uncorrelated across both time and space.}. In the cooperative phase, each user transmits $k_2$ parity-check packets consisting of nonbinary linear combinations of its $k_1$ own information packets and the $k_1(M-1)$ partners' information packets (if decoded correctly). If a user cannot correctly decode an information packet from one of its partners, this information packet is replaced by an all-zero packet in the formation of the linear combination. Herein we assume that, as in~\cite{xiao.09}, the BS knows how each parity packet is formed. This may lead to some overhead, however, if $I_j(t)$ is long enough, the overhead is negligible. Thus, the GDNC overall rate is given by:
\begin{equation} \label{eq:new_rate}
R_{\text{GDNC}} = \frac{k_1M}{k_1M+k_2M} = \frac{k_1}{k_1+k_2}.
\end{equation}
From (\ref{eq:singleton}) and (\ref{eq:new_rate}), we can see that the diversity order of the GDNC scheme is upper bounded by
\begin{equation} \label{eq:UB_GDNC}
D_{\text{GDNC}} \leq k_2M+1.
\end{equation}
However, due to the errors in inter-user channels, this upper bound cannot be achieved. In the next section, we find the achievable diversity order of the proposed scheme.

\section{Outage Probability and Diversity Order} \label{sec:outage}

Denote by $D_{j,t} \subseteq  \{ 1, \ldots, M \}$ the index set corresponding to the users that correctly decoded $I_j(t)$, the information packet of user $j$ in time slot $t$ in the broadcast phase. For convenience, include index $j$ itself to $D_{j,t}$. The number of users in $D_{j,t}$ is denoted by $|D_{j,t}|$. Let $P_e$ again be the outage probability of a single channel, and let $\overline{D}_{j,t}$ denote the complement set $\{ 1, \ldots, M \} \backslash D_{j,t}$, {\em i.e.}, $\overline{D}_{j,t}$ contains the indices of the users which could not decode $I_j(t)$ correctly. The probability of $\overline{D}_{j,t}$ is approximately $P_e^{|\overline{D}_{j,t}|}$.

We define a new set $\mathcal{D}_{j,t}(I)$ as the set of all messages correctly decoded by the users in $D_{j,t}$ in the broadcast phase, including $I_{j}(t)$.  There are at least $|\mathcal{D}_{j,t}(I)|+|D_{j,t}|k_2$ packets containing messages of $\mathcal{D}_{j,t}(I)$ (the $|\mathcal{D}_{j,t}(I)|$ messages themselves from the systematic part and $|D_{j,t}|k_2$ parities). Assuming that a well designed network code is used, the $|\mathcal{D}_{j,t}(I)|$ messages can be recovered from any $|\mathcal{D}_{j,t}(I)|$ received packets\footnote{For the moment being, we are just assuming the existence of such codes. We will prove their existence later.}. For a fixed $D_{j,t}$ (which fixes $\overline{D}_{j,t}$ as well), an outage for $I_j(t)$ is declared only if the direct transmission $I_j(t)$ and at least $|D_{j,t}|k_2$ out of the remaining $|\mathcal{D}_{j,t}(I)|+|D_{j,t}|k_2-1$ received packets are in outage, which occurs with probability
\begin{subequations} \label{eq:outage}
\begin{align}
P_{o,j}(\overline{D}_{j,t}) &= P_e\biggl[\overbrace{{{|\mathcal{D}_{j,t}(I)|+|D_{j,t}|k_2-1}\choose{|D_{j,t}|k_2}}\frac{P_e^{|D_{j,t}|k_2}}{(1-P_e)^{1-|\mathcal{D}_{j,t}(I)|}}}^{|D_{j,t}|k_2 \text{ in outage}}+ \ldots \nonumber \\
&\quad \ldots+\underbrace{{{|\mathcal{D}_{j,t}(I)|+|D_{j,t}|k_2-1}\choose{|\mathcal{D}_{j,t}(I)|+|D_{j,t}|k_2-1}}\frac{P_e^{|\mathcal{D}_{j,t}(I)|+|D_{j,t}|k_2-1}}{(1-P_e)^0}}_{\text{All the }|\mathcal{D}_{j,t}(I)|+|D_{j,t}|k_2-1 \text{ in outage}} \biggr] \label{eq:outagea} \\
&\approx P_e \left[{{|\mathcal{D}_{j,t}(I)|+|D_{j,t}|k_2-1}\choose{|D_{j,t}|k_2}} P_e^{|D_{j,t}|k_2}\right] \label{eq:outageb} \\
&= \gamma(k_1,k_2,\overline{D}_{j,t}) P_e^{(M-|\overline{D}_{j,t}|)k_2+1}, \label{eq:outagec}
\end{align}
\end{subequations}
%--------------------------------------------------------------------------------------------------

%\begin{subequations} \label{eq:outage}
%\begin{align}
%P_{o,j}(\overline{D}_{j,t}) &= P_e\biggl[\overbrace{{{|\mathcal{D}_{j,t}(I)|+|D_{j,t}|k_2-1}\choose{|D_{j,t}|k_2}}P_e^{|D_{j,t}|k_2}(1-P_e)^{|\mathcal{D}_{j,t}(I)|-1}}^{|D_{j,t}|k_2 \text{ in outage}}+ \ldots \nonumber \\
%&\quad \ldots+\underbrace{{{|\mathcal{D}_{j,t}(I)|+|D_{j,t}|k_2-1}\choose{|\mathcal{D}_{j,t}(I)|+|D_{j,t}|k_2-1}}P_e^{|\mathcal{D}_{j,t}(I)|+|D_{j,t}|k_2-1}(1-P_e)^0}_{\text{All the }|\mathcal{D}_{j,t}(I)|+|D_{j,t}|k_2-1 \text{ in outage}} \biggr] \label{eq:outagea} \\
%&\approx P_e \left[{{|\mathcal{D}_{j,t}(I)|+|D_{j,t}|k_2-1}\choose{|D_{j,t}|k_2}} P_e^{|D_{j,t}|k_2}\right] \label{eq:outageb} \\
%&= \gamma(k_1,k_2,\overline{D}_{j,t}) P_e^{(M-|\overline{D}_{j,t}|)k_2+1}, \label{eq:outagec}
%\end{align}
%\end{subequations}

%-------------------------------------------------------------------------------------
where ${{n}\choose{k}}$ is the binomial coefficient and \[\gamma(k_1,k_2,\overline{D}_{j,t})={{|\mathcal{D}_{j,t}(I)|+|D_{j,t}|k_2-1}\choose{|D_{j,t}|k_2}}\] is a positive integer representing the number (multiplicity) of outage patterns leading to that same probability. We can see that $\gamma(k_1,k_2,\overline{D}_{j,t})$ increases as $|\mathcal{D}_{j,t}(I)|$ increases. In particular, since $k_1\leq |\mathcal{D}_{j,t}(I)|\leq Mk_1$, we must have that
%\begin{equation} \label{eq:upper_gamma}
%{{k_1+k_2-1}\choose{k_2}} \leq \gamma(k_1,k_2,\{1,\ldots,M \} \backslash \{j \}) \leq {{Mk_1+k_2-1}\choose{k_2}}.
%\end{equation}

% ---- 2 column version ---------------
\begin{equation} \label{eq:upper_gamma}
\! \! {{k_1 \! + \! k_2 \!- \!1}\choose{k_2}} \!\!  \leq \! \gamma(k_1,k_2,\!  \{1,\!  \ldots \! ,M \}\!  \backslash \!  \{j \}\! ) \! \leq \! {{Mk_1\!+ \! k_2 \! - \! 1}\choose{k_2}}.
\end{equation}
% ---- 2 column version ---------------
It should be noted that \eqref{eq:outage} is the worst case, since the messages in $\mathcal{D}_{j,t}(I)$ (except $I_{j}(t)$) might also be decoded by some users in $\overline{D}_{j,t}$, reducing the outage probability of the messages in $|\mathcal{D}_{j,t}(I)|$.

The outage probability is then given by
\begin{subequations} \label{eq:over_prob}
\begin{align}
P_{o,j} & =  \sum_{\overline{D}_{j,t}} P_e^{|\overline{D}_{j,t}|}(1-P_e)^{(M-1)-|\overline{D}_{j,t}|} P_{o,j}(\overline{D}_{j,t})  \label{eq:newprobaa}\\
& \approx  \sum_{\overline{D}_{j,t}}  P_e^{(M-|\overline{D}_{j,t}|)k_2+|\overline{D}_{j,t}|+1} \gamma(k_1,k_2,\overline{D}_{j,t}) \label{eq:newproba}\\
& \approx  {{M-1}\choose{|\overline{D}_{j,t}|^*}} P_e^{(M-|\overline{D}_{j,t}|^*)k_2+|\overline{D}_{j,t}|^*+1}\gamma(k_1,k_2,\overline{D}_{j,t}) \label{eq:newprobb} \\
& =  \gamma(k_1,k_2,\{1,\ldots,M \} \backslash \{j \}) P_e^{M+k_2}, \label{eq:newprobc}
\end{align}
\end{subequations}
where $P_e^{|\overline{D}_{j,t}|}(1-P_e)^{(M-1)-|\overline{D}_{j,t}|}$ is the probability of $|\overline{D}_{j,t}|$ out of $M-1$ inter-user channels in time slot $t$ being in outage, and $|\overline{D}_{j,t}|^*$ corresponds to the $|\overline{D}_{j,t}|$ value that results in the lowest exponent term in \eqref{eq:newproba}, which, for $k_2 \ge 2$, is $|\overline{D}_{j,t}|^*=M-1$. For $k_2=1$, the exponent in \eqref{eq:newproba} is equal to $M+k_2$ independently of $|\overline{D}_{j,t}|$. We have proved the following result.

\begin{theorem} \label{th:new_diversity}
The diversity order of the GDNC scheme for an appropriately designed network code with sufficiently large field size is $D_{\text{GDNC}}=M+k_2$.
\end{theorem}

We can see that, when $k_1=1$ and $k_2=M-1$, the proposed scheme reduces to the DNC scheme~\cite{xiao.09,xiao.10}, with rate $k_1/(k_1+k_2) = 1/M$ and diversity order $M+k_2 = 2M-1$, as mentioned at the end of Section \ref{subsec:xiao}. In particular, for $k_1 = k_2 = 1$ we have the 2-user DNC scheme shown in Fig.~\ref{fig:Xiao}. The design approach in the current paper allows for different values of $k_1$ and $k_2$, effecting different rates and different diversity orders. By varying $k_1$ and $k_2$ independently, we can obtain a tradeoff between rate and diversity. From \eqref{eq:new_rate} and Theorem \ref{th:new_diversity}, we can see that an appropriate choice of $k_1$ and $k_2$ can simultaneously improve rate and diversity order over the DNC scheme.

%According to \eqref{eq:upper_gamma}, we can upper bound the outage probability by:
%\begin{equation}
%P_{o,j} \leq {{Mk_1+k_2-1}\choose{k_2}} P_e^{M+k_2}.
%\end{equation}

\section{On the Network Code Design} \label{sec:net_design}

 It was shown in Theorem \ref{th:new_diversity} that the proposed GDNC scheme can achieve diversity order $M+k_2$ if the network code is appropriately designed. In this section, we design such network codes in the light of classical coding theory. First, we present some useful linear block codes properties.

Let $\mathcal{C}$ be an $(n,k,d_{min})$ linear block code over GF$(q)$ with a systematic generator matrix $\textbf{G}$ given by
\begin{eqnarray} \label{eq:G}
\textbf{G} &=&
\left[
\begin{array}{cccc}
\textbf{g}_1 & \textbf{g}_2 & \cdots  &\textbf{g}_k \\
\end{array} \right]^T \nonumber \\
&=&
\left[
\begin{array}{cccc|cccc}
1       & 0     & \cdots & 0    & p_{1,1}   & p_{1,2}   & \cdots & p_{1,n-k}   \\
0       & 1     & \cdots & 0    & p_{2,1}   & p_{2,2}   & \cdots & p_{2,n-k}   \\
\vdots  &\vdots & \ddots &\vdots& \vdots    &\vdots     & \ddots & \vdots      \\
0       & 0     & \cdots & 1    & p_{k,1}& p_{k,2}& \cdots & p_{k,n-k}\\
\end{array} \right] \nonumber \\
&=& \left[
\begin{array}{c|c}
\textbf{I}_{k}  & \textbf{P}_{k \times n-k}
\end{array} \right].
\end{eqnarray}

Recall from~\eqref{eq:singleton} that the minimum distance of $\mathcal{C}$ is upper bounded by $d_{min} \leq n-k+1$. From now on, we assume that code $\mathcal{C}$ is MDS, {\it i.e.}, that $d_{min} = n-k+1$. We will make use of the following results regarding the systematic generator matrix of MDS codes.
\begin{theorem}\emph{(\cite[Chap.11, Th.8]{sloane.77})} \label{th:nonsingular}
An $(n,k,d_{min})$ code $\mathcal{C}$ with generator matrix $\emph{G}=[\textbf{I}|\textbf{P}]$, as in~\eqref{eq:G}, is MDS if and only if every square submatrix %(formed from any $i$ rows and any $i$ columns, for any $i=1,2,\ldots,\min\{n,n-k\}$)
of $\textbf{P}$ is nonsingular.
\end{theorem}
\noindent
The following lemma is a direct consequence of Theorem \ref{th:nonsingular}.
\begin{lemma} \label{le:Pfull}
For any $(n,k,d_{min})$ MDS code $\mathcal{C}$ with generator matrix $\emph{G}=[\textbf{I}|\textbf{P}]$, the matrix $\textbf{P}$ has no zero entry.
\end{lemma}
\begin{IEEEproof}
From Theorem \ref{th:nonsingular}, any one-by-one submatrix of $\textbf{P}$ is nonsingular for MDS codes. Alternatively, any row of $\textbf{G}$ is a codeword, and must have Hamming weight at least $n-k+1$.
\end{IEEEproof}

\subsection{A Theory of Faulty Generator Matrices} \label{subsec:faulty}

In this section, we study some properties of new block codes obtained from systematic MDS codes by zeroing some entries of its generator matrix, and we refer to the obtained generator matrix as a {\it faulty generator matrix}. In particular, we are interested in the minimum Hamming distance or the least erasure-correcting capability of MDS codes that can be assured when these codes are encoded by a faulty generator matrix. We will see that the theory of faulty generator matrices we develop next is related to and will shed light on our main problem: the design of network codes to maximize the diversity order of cooperative wireless communications assuming possible link failures.

When a set of entries of the generator matrix $\textbf{P}$ of the MDS code $\mathcal{C}$ is  replaced by zeros, the new code $\mathcal{C}'$  produced is no longer MDS. Let $A=\{(a_1,b_1),(a_2,b_2),...\}$ be a subset of entries of $\textbf{P}$ that are replaced by zeros, where $(a,b)$  is the entry in row $a$ and column $b$ of matrix $\textbf{P}$. We call $A$ a {\it fault}. %The act of replace a set of elements of $\textbf{P}$ by zero elements is here called as a \emph{fault}.
Let $\mathcal{A} = \{A_0, A_1, \ldots , A_{f-1}\}$ be a collection of $f$ faults. We consider that two different faults $A_i$ and $A_j$ cannot contain a common entry of the matrix $\textbf{P}$. That is, $A_i \cap A_j = \emptyset$. It is also considered that every fault has fixed cardinality, {\it i.e.}, $|A_i|=|A| \; \forall i$. Although these considerations may look too restrictive, we will see that they are sufficient for the purposes of solving our main problem.

Let $\chi=(\chi_0,\ldots,\chi_{f-1})$ be the binary indicator vector associated with the occurrence of faults, where
 \[\begin{cases} \chi_i =1& \text{if $A_i$ occurs}\\ \chi_i=0 & \text{if $A_i$ does not occur.} \end{cases} \]
For a nonnegative integer $i$, let $b(i)$ denote the binary (vector) representation of $i$. We denote the collection of all possible combinations of faults by $\mathcal{B} = \{ B_{b(0)}, B_{b(1)}, ... , B_{b(2^f - 1)} \}$, with $B_\chi= \{(a,b)|(a,b) \in \bigcup_{i:\chi_i=1} A_i\}$. Each event $B_\chi$, which consists of the occurrence of $|B_\chi|/|A|$ faults, where $|\cdot|$ stands for the cardinality of the set, gives rise to a new generator matrix of a block code $\mathcal{C}(B_\chi)$ with minimum Hamming distance $d_{min}(\mathcal{C}(B_\chi))$. We also define the {\em minimum composite distance} of the code $\mathcal{C}(B_\chi)$ as
\begin{equation} \label{eq:dmineffBx}
d_{min}^{\text{comp}}(\mathcal{C}(B_\chi)) \triangleq d_{min}(\mathcal{C}(B_\chi)) + |B_\chi|/|A| ,
\end{equation}
which is composed of its minimum Hamming distance $d_{min}(\mathcal{C}(B_\chi))$ plus a ``compensation'' term related to the number of faults in the combination $B_\chi$.

A parameter of fundamental importance to indicate the performance of a MDS code subject to a set of faulty generator matrices is the least minimum composite distance of any possible combination of faults. In particular, given a MDS code $\mathcal{C}$ and a collection of $f$ faults $\mathcal{A}$, this distance is defined as:
\begin{equation} \label{eq:dmineff}
d_{min}^{\text{comp}}(\mathcal{C},\mathcal{A}) \triangleq \min_{B_\chi \in \mathcal{B}} \left\{ d_{min}^{\text{comp}}(\mathcal{C}(B_\chi)) \right\},
\end{equation}
where, when there is no confusion, it is simply called {\em minimum composite distance}.

We next prove a sequence of results which constitute a theory of faulty generator matrices.

\begin{lemma} \label{lem:triangle2}
The weight between two codewords $y$ and $z$ is lower bounded by the difference between their individual weights, {\em i.e.}, $w(y\boxplus z)\geq w(y)-w(z)$.
\end{lemma}
\begin{IEEEproof}
From the triangular inequality, $w(y' \boxplus z')\leq w(y')+w(z')$. If we set $y'=y\boxplus z$ and $z'=\boxminus z$, where $\boxminus z$ is the additive inverse of $z$, we obtain $w(y\boxplus z)\geq w(y)-w(z)$, which proves the lemma.
\end{IEEEproof}
\begin{theorem} \label{th:k2zeros}
 Let $\mathcal{C}$ be an $(n,k,d_{min})$ MDS code with systematic generator matrix $\textbf{G}=[\textbf{I}|\textbf{P}]$. The replacement by zeros of $\delta$ entries in any single row of the matrix $\textbf{P}$ gives rise to an $(n,k,d_{min}-\delta)$ code $\mathcal{C}'$.
\end{theorem}
The proof is provided in Appendix~\ref{ap:k2zeros}.

\begin{lemma} \label{le:0column}
Let $\mathcal{C}'$ be an $(n,k,n-k-\delta+1)$ code (with generator matrix $\textbf{G}'=[\textbf{I}|\textbf{P}']$) obtained from an $(n,k,n-k+1)$ MDS code $\mathcal{C}$ according to Theorem \ref{th:k2zeros}. The replacement by zeros of additional entries in any of the $\delta$ columns of $\textbf{P}'$ where zeros have previously been inserted gives rise to a code $\mathcal{C}''$ with the same minimum Hamming distance as $\mathcal{C}'$.
\end{lemma}
\begin{IEEEproof}
Each codeword of $\mathcal{C}''$ can be obtained from a codeword of $\mathcal{C}$ by changing $\delta$ positions, which are fixed for all codewords. Therefore, we must have that $d_{min}(\mathcal{C}'') \ge d_{min}(\mathcal{C})-\delta$. Since $\textbf{g}'_{i^*}$ (see Appendix~\ref{ap:k2zeros}), a codeword of $\mathcal{C}'$ of Hamming weight $n-k-\delta+1$, is also a codeword of $\mathcal{C}''$, the Lemma is proved.
\end{IEEEproof}

\begin{lemma} \label{le:lb}
If $\mathcal{C}$ is an $(n,k,d_{min})$ MDS code with systematic generator matrix $\textbf{G}=[\textbf{I}|\textbf{P}]$, the replacement by zeros of $\delta$ arbitrary entries of its matrix $\textbf{P}$ generates a code $\mathcal{C}'$ with minimum Hamming distance $d_{min}(\mathcal{C}') \geq d_{min}-\delta$.
\end{lemma}
\begin{IEEEproof}
Similar to Lemma \ref{le:0column}, with equality if the new code $\mathcal{C}'$  has a codeword with Hamming weight $d_{min}-\delta$. For example, if all the $\delta$ replacements occur in a single row of $\textbf{P}$.
\end{IEEEproof}

According to Lemma~\ref{le:lb}, a single fault $A$ (confined to the matrix $\textbf{P}$) can reduce the minimum Hamming distance of the original MDS code by at most $|A|$. On the other hand, Theorem \ref{th:k2zeros} tells us that the worst scenario for a single fault $A$ is when all of its entries belong to a single row of $\textbf{P}$. The last result of this section refers to the worst scenario for a collection of faults.

\begin{lemma} \label{le:dagg}
Let $\mathcal{C}$ be an $(n,k,d_{min})$ MDS code with systematic generator matrix $\textbf{G}=[\textbf{I}|\textbf{P}]$. Consider a collection $\mathcal{A} = \{A_0, A_1, \ldots , A_{f-1}\}$ of $f$ faults, where $|A_i|=|A| \geq 1, \; \forall i$. If all the faults occur in a single row of $\textbf{P}$, then the minimum composite distance of $\mathcal{C}$ subject to $\mathcal{A}$ is a strictly decreasing function of the number of faults.
\end{lemma}
\begin{IEEEproof}
From Theorem \ref{th:k2zeros}, each such a fault reduces the minimum Hamming distance by exactly $|A|\geq 1$, and contributes with only 1 to the ``compensation'' term of the  minimum composite distance. In the special case where $|A|=1$, the minimum composite distance of $\mathcal{C}$ is constant independently of the number of faults.
\end{IEEEproof}

\subsection{Maximum-Diversity GDNC} \label{subsec:max_div_gdnc}

%In cooperative wireless communications, the diversity order is related to the minimum number of correctly received packets (or symbols) at the BS with which the information packets from all users can be recovered. From the coding theory point of view, in terms of block coding, this is equivalent to the erasure correction capability of the block code. It is well-known that the transmitted codeword of a linear block code with minimum distance $d_{min}$ can be recovered if up to $d_{min}-1$ of its positions have been erasured by the channel~\cite[Chapter 3]{costello.04}. The connection between these two problems establishes that the diversity order of a $M$-user system with rate $k/n$, under the assumption of perfect inter-user channels, is equal to the minimum Hamming distance of the $(n,k)$ equivalent block code.

%According to \eqref{eq:new_rate}, the $M$-user GDNC scheme has overall rate $R_{\text{GDNC}} = \frac{k_1M}{k_1M+k_2M}$. For perfect inter-user channels, the bound presented in \eqref{eq:singleton} states that the GDNC's diversity is upper bounded by $k_2M+1$.
%
%However, when inter-user channel errors are taken into account, it was shown in Theorem \ref{eq:over_prob} that the diversity order of the GDNC scheme for sufficiently large field size is $D_{\text{GDNC}}=M+k_2$.

%Yet, that diversity order is only achieved if the code is properly designed. Assuming that, as in \cite{xiao.09}, the base station knows how each receiving network codeword is formed,

Recall that our main problem is to design a linear network code for the GDNC scheme that, given all the inter-user error patterns and taking into account their probabilities of occurrence, assures that the diversity order is $M + k_2$. The solution to this problem is our main result, presented next.

\begin{theorem} \label{th:gndc_mds}
If a systematic generator matrix of a MDS code $\mathcal{C}$ with minimum Hamming distance $d_{\text{min}} = Mk_2+1$ is used as a transfer matrix of the GDNC scheme, then the diversity order $D_{\text{GDNC}} = M + k_2$ is guaranteed.
 \end{theorem}
The proof of this theorem is provided in Appendix~\ref{ap:gndc_mds}.

\begin{theorem} \label{th:quasi_mds}
If a generator matrix of a non-MDS code is used as the transfer matrix in the GDNC scheme, the maximum diversity order presented in Theorem \ref{th:new_diversity} is not guaranteed.
\end{theorem}
The proof is provided in Appendix~\ref{ap:quasi_mds}.

Now that we have proved that MDS codes are sufficient to achieve the maximum diversity order of the GDNC scheme, the network code design problem is reduced to a MDS code design, without the hassle of testing a large number of network codes, taking into account all the inter-user channel error patterns, and their impact on the system's diversity. The well-known class of MDS codes called Reed-Solomon (RS) codes \cite[Chap.3]{costello.04} \cite[Chap.10]{sloane.77}, for example, may be used. In the destination node, the users' message packets can be recovered with some classical decoder for RS codes, {\em e.g.} \cite{koeter.03b,koeter.PhD.00}.

The generator matrix of an $(n,k,n-k+1)$ RS (or extended RS) code over GF$(q)$ in systematic form can be easily obtained with the software application SAGE~\cite{sage}, entering the following code:

\begin{center}
\fbox{
\begin{minipage}[c]{5.5cm}
{\tt
F.<alpha> = GF(q) \newline
C = ReedSolomonCode(n,k,F) \newline
G = C.gen\_mat() \newline
show(G.echelon\_form())}
\end{minipage}}
\end{center}
It should be noted that the field size $q$ must be sufficiently large~\cite[Chap.11, Coroll.7]{sloane.77}. In this regard, we have the following result.
\begin{theorem} \label{th:q_sufficient}
There exists an MDS code $(n=M(k_1+k_2), k=Mk_1, d_{min} = Mk_2 + 1)$ over GF($q$) (where $q$ is a power of prime) that can serve a GDNC system achieving maximum diversity order $M + k_2$ if $q \ge M(k_1 + k_2)$.
\end{theorem}
The proof is provided in the Appendix~\ref{ap:q_sufficient}.

We can now compare the field size necessary for the full diversity order to be achieved in the GDNC scheme (according to the bound presented in Theorem~\ref{th:q_sufficient}) and the simplified DNC scheme (as proved in~\cite{xiao.10} and shown in~\eqref{eq:q_dnc}). This comparison is exemplified in Table~\ref{tab:field_size}.

\begin{table}[!htb]
\centering
\caption{Field size necessary for the diversity order achieved by the schemes GDNC (with $k_1=1$ e $k_2=M-1$) and DNC to be guaranteed.}
\label{tab:field_size}
\begin{IEEEeqnarraybox}[\IEEEeqnarraystrutmode\IEEEeqnarraystrutsizeadd{2pt}{1pt}]{v/c/v/c/v/c/v/c/v/c/v}
\IEEEeqnarrayrulerow\\
&M&& 2 &&3 &&4 && 5 & \\
\IEEEeqnarrayrulerow\\
\IEEEeqnarrayseprow[3pt]\\
& q_{\min} (\text{GDNC}) && 4 && 9 && 16 && 25 & \\
\IEEEeqnarrayseprow[3pt]\\
\IEEEeqnarrayrulerow\\
& q_{\min} (\text{DNC})&& 3 && 28 && 455 && 10626 & \\
\IEEEeqnarrayseprow[3pt]\\
\IEEEeqnarrayrulerow
\end{IEEEeqnarraybox}
\end{table}

We emphasize that the values shown in Table~\ref{tab:field_size} regarding the DNC scheme correspond to the finite field size values necessary for the diversity order $2M-1$ to be achieved, according to~\eqref{eq:q_dnc}. However, since the GDNC scheme with parameters $k_1=1$ and $k_2=M-1$ reduces to the DNC scheme, we can see that our result in Theorem~\ref{th:q_sufficient} represents a much tighter lower bound on the field size.
%\textcolor{red}{Olhar \cite[Chap.11, Th.9]{sloane.77}!}

Tables \ref{tab:RS_codesM2} and \ref{tab:RS_codesM3} list the parity matrix $\textbf{P}$ described in \eqref{eq:G} of some  MDS codes obtained from RS codes (as explained in the proof of Theorem \ref{th:q_sufficient}) which can be used as the GDNC's network transfer matrix, for $M=2$ and $M=3$ users, respectively. These matrices were obtained using SAGE~\cite{sage}.

\begin{table}[!htb]
\centering
\caption{Network codes obtained from RS codes for $M=2$ users.}
\label{tab:RS_codesM2}
\begin{IEEEeqnarraybox}[\IEEEeqnarraystrutmode\IEEEeqnarraystrutsizeadd{2pt}{1pt}]{v/c/v/c/v/c/v/c/v/c/v}
\IEEEeqnarrayrulerow\\
&k_1&&k_2&&\text{Code rate}&&\text{Field size }q&& \text{Parity matrix } \textbf{P}& \\
\IEEEeqnarraydblrulerow\\ \IEEEeqnarrayseprow[3pt]\\
%&&&&&\text{Rate}&&\mbox{size }q&&&\\ \IEEEeqnarraydblrulerow\\ \IEEEeqnarrayseprow[3pt]\\
&1&&1&&2/4&&4&&\left[\begin{IEEEeqnarraybox*}[][c]{,c/c,}
3 & 2 \\
2 & 3%
\end{IEEEeqnarraybox*}\right]&\IEEEeqnarraystrutsize{0pt}{0pt}\\
\IEEEeqnarrayseprow[3pt]\\
\IEEEeqnarrayrulerow\\
\IEEEeqnarrayseprow[3pt]\\
&1&&2&&2/6&&8&&\left[\begin{IEEEeqnarraybox*}[][c]{,c/c/c/c,}
3 & 5 & 1 & 7 \\
2 & 4 & 3 & 6%
\end{IEEEeqnarraybox*}\right]&\IEEEeqnarraystrutsize{0pt}{0pt}\\
\IEEEeqnarrayseprow[3pt]\\
\IEEEeqnarrayrulerow\\
\IEEEeqnarrayseprow[3pt]\\
&\raisebox{-5pt}[0pt][0pt]{$2$}&&\raisebox{-5pt}[0pt][0pt]{$1$}&&\raisebox{-5pt}[0pt][0pt]{$4/6$}&&\raisebox{-5pt}[0pt][0pt]{$8$}
&&\text{Obtained from the $4/8$ code by}&\IEEEeqnarraystrutsize{0pt}{0pt}\\
\IEEEeqnarrayseprow[2pt]\\
&&&&&&&&&\text{puncturing the last 2 columns}&\IEEEeqnarraystrutsize{0pt}{0pt}\\
\IEEEeqnarrayseprow[3pt]\\
\IEEEeqnarrayrulerow\\
\IEEEeqnarrayseprow[3pt]\\
&2&&2&&4/8&&8&&\left[\begin{IEEEeqnarraybox*}[][c]{,c/c/c/c,}
3 & 7 & 3 & 6 \\
5 & 7 & 7 & 4 \\
2 & 4 & 6 & 1\\
5 & 5 & 3 & 2%
\end{IEEEeqnarraybox*}\right]&\IEEEeqnarraystrutsize{0pt}{0pt}\\
\IEEEeqnarrayseprow[3pt]\\
\IEEEeqnarrayrulerow\\
\IEEEeqnarrayseprow[3pt]\\
&\raisebox{-5pt}[0pt][0pt]{$2$}&&\raisebox{-5pt}[0pt][0pt]{$3$}&&\raisebox{-5pt}[0pt][0pt]{$4/10$}&&\raisebox{-5pt}[0pt][0pt]{$16$}
&&\text{Obtained from the $4/12$ code by}&\IEEEeqnarraystrutsize{0pt}{0pt}\\
\IEEEeqnarrayseprow[2pt]\\
&&&&&&&&&\text{puncturing the last 2 columns}&\IEEEeqnarraystrutsize{0pt}{0pt}\\
\IEEEeqnarrayseprow[3pt]\\
\IEEEeqnarrayrulerow\\
\IEEEeqnarrayseprow[3pt]\\
&2&&4&&4/12&&16&&\left[\begin{IEEEeqnarraybox*}[][c]{,c/c/c/c/c/c/c/c,}
14 & 5 & 7  & 10 & 4 & 10 & 8 & 9 \\
12 & 4 & 14 & 5  & 12& 12 & 14& 7\\
13 & 6 & 12 & 15 & 8 & 1 & 4 & 10\\
14 & 6 & 4 & 1 & 1  & 6 & 3 & 5%
\end{IEEEeqnarraybox*}\right]&\IEEEeqnarraystrutsize{0pt}{0pt}\\
\IEEEeqnarrayseprow[3pt]\\
\IEEEeqnarrayrulerow\\
\IEEEeqnarrayseprow[3pt]\\
&\raisebox{-5pt}[0pt][0pt]{$3$}&&\raisebox{-5pt}[0pt][0pt]{$2$}&&\raisebox{-5pt}[0pt][0pt]{$6/10$}&&\raisebox{-5pt}[0pt][0pt]{$16$}
&&\text{Obtained from the $6/12$ code by}&\IEEEeqnarraystrutsize{0pt}{0pt}\\
\IEEEeqnarrayseprow[2pt]\\
&&&&&&&&&\text{puncturing the last 2 columns}&\IEEEeqnarraystrutsize{0pt}{0pt}\\
\IEEEeqnarrayseprow[3pt]\\
\IEEEeqnarrayrulerow\\
\IEEEeqnarrayseprow[3pt]\\
&3&&3&&6/12&&16&&\left[\begin{IEEEeqnarraybox*}[][c]{,c/c/c/c/c/c,}
11 & 2 & 4  & 6  & 14 & 12  \\
1 & 11 & 13 & 10 & 14 & 10\\
2 & 4 & 2   & 10 & 5  & 9\\
6 & 13 & 12 & 11 & 8  & 12\\
4 & 12 & 12 & 2  & 6  & 6\\
11 & 13 & 10& 14 & 10 & 4%
\end{IEEEeqnarraybox*}\right]&\IEEEeqnarraystrutsize{0pt}{0pt}\\
\IEEEeqnarrayseprow[3pt]\\
\IEEEeqnarrayrulerow
\end{IEEEeqnarraybox}
\end{table}

\begin{table}[!htb]
\centering
\caption{Network codes obtained from RS codes for $M=3$ users.}
\label{tab:RS_codesM3}
\begin{IEEEeqnarraybox}[\IEEEeqnarraystrutmode\IEEEeqnarraystrutsizeadd{2pt}{1pt}]{v/c/v/c/v/c/v/c/v/c/v}
\IEEEeqnarrayrulerow\\
&k_1&&k_2&&\text{Code rate}&&\text{Field size }q&& \text{Parity matrix } \textbf{P}& \\
\IEEEeqnarraydblrulerow\\ \IEEEeqnarrayseprow[3pt]\\
& 1 && 1 && 3/6 && 8 &&\left[\begin{IEEEeqnarraybox*}[][c]{,c/c/c,}
4 & 1 & 5 \\
3 & 1 & 3\\
6 & 1 & 7%
\end{IEEEeqnarraybox*}\right]&\IEEEeqnarraystrutsize{0pt}{0pt}\\
\IEEEeqnarrayseprow[3pt]\\
\IEEEeqnarrayrulerow\\
\IEEEeqnarrayseprow[3pt]\\
&\raisebox{-5pt}[0pt][0pt]{$2$}&&\raisebox{-5pt}[0pt][0pt]{$1$}&&\raisebox{-5pt}[0pt][0pt]{$6/9$}&&\raisebox{-5pt}[0pt][0pt]{$16$}
&&\text{Obtained from the $6/12$ code by}&\IEEEeqnarraystrutsize{0pt}{0pt}\\
\IEEEeqnarrayseprow[2pt]\\
&&&&&&&&&\text{puncturing the last 3 columns}&\IEEEeqnarraystrutsize{0pt}{0pt}\\
\IEEEeqnarrayseprow[3pt]\\
\IEEEeqnarrayrulerow\\
\IEEEeqnarrayseprow[3pt]\\
& 1 && 2 && 3/9 && 16 &&\left[\begin{IEEEeqnarraybox*}[][c]{,c/c/c/c/c/c,}
15& 11& 1 & 14 & 5  & 11\\
8 & 5 & 1 & 8  & 13 & 4\\
6 & 15& 1 & 7  & 9  & 14%
\end{IEEEeqnarraybox*}\right]&\IEEEeqnarraystrutsize{0pt}{0pt}\\
\IEEEeqnarrayseprow[3pt]\\
\IEEEeqnarrayrulerow\\
\IEEEeqnarrayseprow[3pt]\\
& 2 && 2 && 6/12 && 16 &&\left[\begin{IEEEeqnarraybox*}[][c]{,c/c/c/c/c/c,}
11& 2 & 4  & 6  & 14 & 12 \\
1 & 11& 13 & 10 & 14 & 10  \\
2 & 4 & 2  & 10 & 5  & 9 \\
6 & 13& 12 & 11 & 8  & 12 \\
4 & 12& 12 & 2  & 6  & 6  \\
11& 13& 10 & 14 & 10 & 4 %
\end{IEEEeqnarraybox*}\right]&\IEEEeqnarraystrutsize{0pt}{0pt}\\
\IEEEeqnarrayseprow[3pt]\\
\IEEEeqnarrayrulerow
\end{IEEEeqnarraybox}
\end{table}

\section{Simulation Results} \label{sec:simulations}

In order to support the outage probability analysis of the previous section, we have performed some computer simulations. The frame error rate (FER) was simulated and plotted against the SNR. In our simulations, we adopt $r = 0.5$ bits per channel use\footnote{It should be noted that the choice of the information rate $r$ is irrelevant for the purpose of obtaining the diversity order.}. We assume that there exists a channel code at the link level with which it is possible to recover the transmitted packet if $|h_{j,i,t}|^2\geq g$. If $|h_{j,i,t}|^2 < g$, an outage in the link $j\rightarrow i$ is declared, and the generator matrix $\textbf{G}$ will have a set of its elements replaced by zeros. The number of erroneous frames is computed based on the rank of the resulting faulty generator matrix. The network codes used in the simulations were obtained with the method presented in Section \ref{sec:net_design}. The analytical outage probabilities obtained in this paper were also plotted (dashed lines).

\begin{figure} [!htb]
\centering
\resizebox{9cm}{!}{
\includegraphics{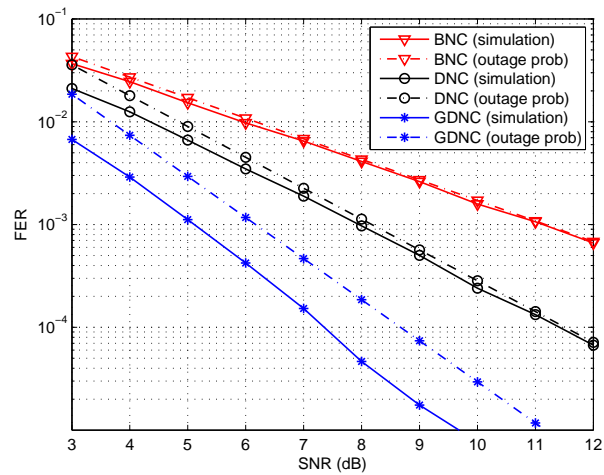}}
\vspace{-0.5cm}
\caption{FER versus SNR (dB) for a 2-user system with rate $R=1/2$, considering the BNC scheme, DNC scheme (in GF(4)) and the proposed GDNC scheme (with $k_1=k_2=2$ and in GF(8), according to Table \ref{tab:RS_codesM2}).}
\label{fig:FER1}
\end{figure}

Fig.~\ref{fig:FER1} shows the FER versus SNR for a 2-user network, considering the BNC scheme~\cite{xiaoL.07}, the DNC scheme~\cite{xiao.09,xiao.10}(Fig.~\ref{fig:Xiao}) over GF(4), and the proposed GDNC with $k_1=k_2=2$ over GF(8), with the generator matrix obtained from Table~\ref{tab:RS_codesM2}. All of them have the same rate equal to 1/2. Regarding the analytical GDNC outage probability, we consider an upper bound which was obtained by substituting the upper bound of $\gamma(k_1,k_2,\overline{D}_{j,t})$ in~\eqref{eq:upper_gamma} into~\eqref{eq:newprobc}. As expected, the proposed scheme achieves a higher diversity order than the two other schemes with the same rate. The SNR gap between the analytical and simulated curves are due to the Gaussian input assumption made in \eqref{eq:mutual_inf}  and the use of the aforementioned upper bound. However, we can see that the simulated diversity order (curve slope) matches the analytical one.
% -------- 2 column version ------------
%\begin{figure} [!htb]
%\centering
%\resizebox{9cm}{!}{
%\includegraphics{FER01.eps}}
%\vspace{-0.5cm}
%\caption{FER versus SNR (dB) for a 2-user system with rate $R=1/2$, considering the BNC scheme, DNC scheme (in GF(4), according to \eqref{eq:G}) and the proposed GDNC scheme (with $k_1=k_2=2$ and in GF(8), according to Table \ref{tab:RS_codesM2}).}
%\label{fig:FER1}
%\end{figure}

It should be mentioned that, since the maximum diversity order 3 of the DNC scheme~\cite{xiao.09,xiao.10} is already achieved with the field GF(4), increasing the field size would not bring any advantages in this case. On the other hand, as discussed in Section~\ref{sec:motivation}, for the proposed GDNC used in our simulations, which corresponds to a rate 4/8 block code, the field size 8 was necessary for $d_{min}=5$. It should also be pointed out that, in the proposed scheme, only the linear combinations are performed over GF($q$) prior to modulation and transmission; the modulation scheme and the transmission mode are irrelevant to the functioning of GDNC, and can be the same for all schemes.

\begin{figure} [!htb]
\centering
\resizebox{9cm}{!}{
\includegraphics{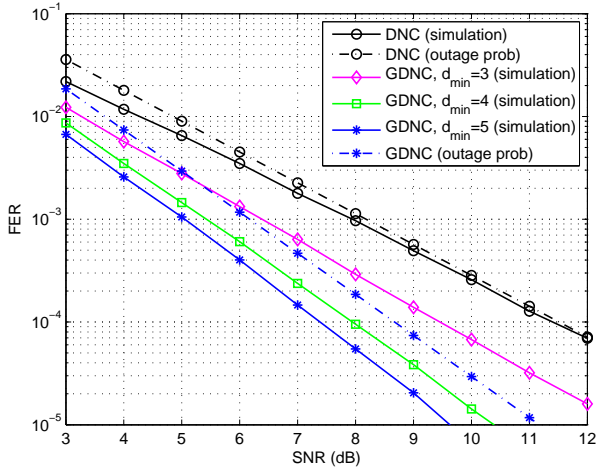}}
\vspace{-0.5cm}
\caption{FER versus SNR (dB) for a 2-user system with rate $R=1/2$, considering the DNC scheme (in GF(4)) and the proposed GDNC scheme (with $k_1=k_2=2$ and in GF(8), according to Table \ref{tab:RS_codesM2} and \ref{tab:Codes_simul}).}
\label{fig:FER2}
\end{figure}

Fig.~\ref{fig:FER2} shows the FER performance for the proposed GDNC scheme illustrated in Fig.~\ref{fig:FER1}, as well as for two other GDNC schemes with non-MDS codes (see Table~\ref{tab:Codes_simul}) with the same parameters $k_1 = k_2 = 2$ and rate $1/2$, but with respective block codes of minimum Hamming distances 3 and 4, whose generator matrices have been randomly generated.

% ---------- 2 column version -----------------
%\begin{figure} [!htb]
%\centering
%\resizebox{9cm}{!}{
%\includegraphics{FER02.eps}}
%\vspace{-0.5cm}
%\caption{FER versus SNR (dB) for a 2-user system with rate $R=1/2$, considering the DNC scheme (in GF(4), according to \eqref{eq:G}) and the proposed GDNC scheme (with $k_1=k_2=2$ and in GF(8), according to Table \ref{tab:RS_codesM2} and \ref{tab:Codes_simul}).}
%\label{fig:FER2}
%\end{figure}

We can see that, with $d_{min}=3$, the diversity order is only 3, while with $d_{min} = 4$ the diversity order is close to the one obtained with the proposed code ($d_{min} = 5$). In this case, a quasi-MDS code achieves the maximum diversity.

\begin{table}[!htb]
\centering
\caption{Network codes obtained from codes with minimum distance 3 and 4.}
\label{tab:Codes_simul}
\begin{IEEEeqnarraybox}[\IEEEeqnarraystrutmode\IEEEeqnarraystrutsizeadd{2pt}{1pt}]{v/c/v/c/v/c/v}
\IEEEeqnarrayrulerow\\
&\text{Fig.}&&d_{\text{min}}&&\text{Parity matrix } \textbf{P} & \\
\IEEEeqnarraydblrulerow\\ \IEEEeqnarrayseprow[3pt]\\
&\text{\ref{fig:FER2} and  \ref{fig:FER3}}&&3&&\left[\begin{IEEEeqnarraybox*}[][c]{,c/c/c/c,}
0 & 5 & 5 & 3 \\
4 & 7 & 7 & 1 \\
7 & 7 & 7 & 1 \\
7 & 2 & 3 & 7%
\end{IEEEeqnarraybox*}\right]&\IEEEeqnarraystrutsize{0pt}{0pt}\\
\IEEEeqnarrayseprow[3pt]\\
\IEEEeqnarrayrulerow\\
\IEEEeqnarrayseprow[3pt]\\
&\text{\ref{fig:FER2}}&&4&&\left[\begin{IEEEeqnarraybox*}[][c]{,c/c/c/c,}
4 & 2 & 5 & 3 \\
5 & 4 & 6 & 4 \\
7 & 6 & 1 & 4 \\
1 & 1 & 3 & 1%
\end{IEEEeqnarraybox*}\right]&\IEEEeqnarraystrutsize{0pt}{0pt}\\
\IEEEeqnarrayseprow[3pt]\\
\IEEEeqnarrayrulerow\\
\IEEEeqnarrayseprow[3pt]\\
&\text{\ref{fig:FER3}}&&4&&\left[\begin{IEEEeqnarraybox*}[][c]{,c/c/c/c,}
0 & 5 & 5 & 5 \\
7 & 4 & 6 & 3 \\
7 & 0 & 4 & 7 \\
2 & 7 & 6 & 0%
\end{IEEEeqnarraybox*}\right]&\IEEEeqnarraystrutsize{0pt}{0pt}\\
\IEEEeqnarrayseprow[3pt]\\
\IEEEeqnarrayrulerow
\end{IEEEeqnarraybox}
\end{table}

\begin{figure} [!htb]
\centering
\resizebox{9cm}{!}{
\includegraphics{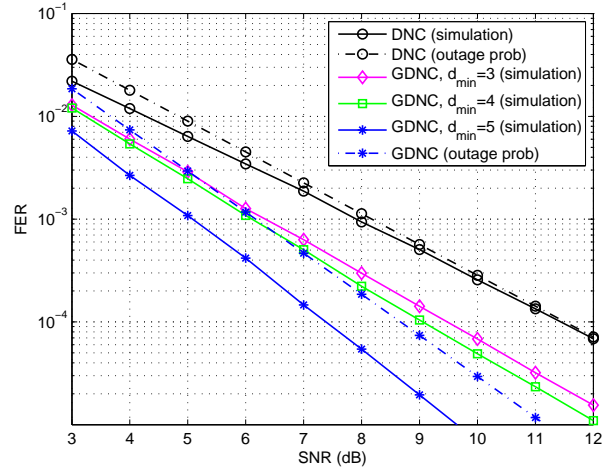}}
\vspace{-0.5cm}
\caption{FER versus SNR (dB) for a 2-user system with rate $R=1/2$, considering DNC scheme (in GF(4)) and the proposed GDNC scheme (with $k_1=k_2=2$ and in GF(8), according to Table \ref{tab:RS_codesM2} and \ref{tab:Codes_simul}).}
\label{fig:FER3}
\end{figure}

However, in Fig.~\ref{fig:FER3}, a different code, also with $d_{min}=4$ (given in Table~\ref{tab:Codes_simul}), is shown to fail to achieve the maximum diversity order. Thus, we can see that by using a generator matrix of a quasi-MDS block code as the network transfer matrix of the GDNC scheme, the diversity order $M+k_2$ is not guaranteed. This is in accordance with Theorem~\ref{th:quasi_mds}.
% --------- 2 column version -------------
%\begin{figure} [!htb]
%\centering
%\resizebox{12cm}{!}{
%\includegraphics{FER03.eps}}
%\vspace{-0.5cm}
%\caption{FER versus SNR (dB) for a 2-user system with rate $R=1/2$, considering DNC scheme (in GF(4), according to \eqref{eq:G}) and the proposed GDNC scheme (with $k_1=k_2=2$ and in GF(8), according to Table \ref{tab:RS_codesM2} and \ref{tab:Codes_simul}).}
%\label{fig:FER3}
%\end{figure}

%In Fig. \ref{fig:FER2} the number of users was chosen to be $M=3$, and the Galois field order was increased to $q=8$. The other parameters are the same as Fig. \ref{fig:FER1}. We can note that the simulation and analytical results still match with good precision.
%\begin{figure} [!hbt]
%\begin{center}
%\resizebox{9.0cm}{!}{
%\includegraphics{SimulM3.eps}}\end{center}
%\caption{Frame Error Rate versus Signal to Noise Ratio (dB) for a system with $M=3$ users, rate $R=1/2$ and Galois field order $q=8$, considering DNC and GDNC schemes (the last one with $k_1=k_2=2$).}
%\label{fig:FER2}
%\end{figure}

%
%It should be noted that the diversity improvement provided by the proposed scheme can be obtained while keeping the same rate as the other schemes and without significant complexity increase, only at the expenses of a larger decoding latency. It is also possible to choose $k_1$ and $k_2$ to simultaneously improve both rate and diversity order. For example, by setting $k_1=4$ and keeping all the other parameters as in Fig. \ref{fig:FER1}, the proposed scheme's rate is increased to $R=2/3$ while its diversity order is still $D=4$.

\section{Conclusions and Final Comments} \label{sec:conclusions}

In this work, we have proposed a generalization of the network coding method in \cite{xiao.09}. The aim of the generalized dynamic-network code (GDNC) is to increase the diversity order of cooperative wireless communications systems without sacrificing the system's rate, what appears to be a drawback of the original method. We have shown that the problem of designing network codes that maximize the diversity order is related to that of designing optimal linear block codes, in the Hamming sense, over a nonbinary finite field.

An outage probability analysis was presented and computer simulations supported the analytical results. The two design parameters $k_1$ and $k_2$ of GDNC may be varied to produce a wide range of rates and diversity orders, and offer a tradeoff between rate and diversity order (when the value of $k_2$ is changed), or even between rate and decoding latency (when $k_1$ is changed), as compared to the baseline system.

Of course, since the GDNC scheme improves the rate-diversity trade-off by coding across more time slots, the decoding complexity and the delay are increased. However, for relatively small number of users and small values of $k_1$ and $k_2$, the decoding complexity and the delay are manageable as efficient decoding methods for RS codes do exist \cite{koeter.03b,koeter.PhD.00}. On the other hand, for a large number of users and/or when the parameters $k_1$ and $k_2$ are very large, the dimensions of the generator matrix of the block code associated with the GDNC scheme may be too large for a practical implementation of the encoders and decoders. Moreover, when the diversity order is already high, a further increase does not bring much additional benefit. In this case, the generator matrix can be made sparse and the overall system can be seen as a generalization of the adaptive network coded cooperation (ANCC) method proposed in \cite{bao.08}. The connection between GDNC and ANCC is currently being investigated.
\appendices

\section{Proof of Theorem \ref{th:k2zeros}} \label{ap:k2zeros}

\begin{IEEEproof}
Let $\textbf{G}=\{\textbf{g}_1,\ldots,\textbf{g}_k\}$ and $\textbf{G}'=\{\textbf{g}'_1,\ldots,\textbf{g}'_k\}$ be the generator matrices of $\mathcal{C}$ and $\mathcal{C}'$, respectively, with $\textbf{g}_i$ and $\textbf{g}'_i$ being the $i$th row of $\textbf{G}$ and $\textbf{G}'$, respectively. Without loss of generality, we assume that $\delta$ elements of the parity part of $\textbf{g}_{i^*}$ are replaced by zeros  for some $i^* \in \{1,\ldots,k\}$. For a nonzero information message $\textbf{u}=\{u_1,\ldots,u_k\}$, the Hamming weight of the corresponding nonzero codeword $\textbf{v}'$ is
%\begin{eqnarray}
%w(\textbf{v}')&=&w(\textbf{uG}') \nonumber \\
%&=&w(u_1 \textbf{g}_1'\boxplus \ldots \boxplus u_{i^*} \textbf{g}_{i^*}' \boxplus \ldots \boxplus u_k \textbf{g}_k') \nonumber \\
%&\overset{\text{(a)}}{=}&w(u_1 \textbf{g}_1 \boxplus \ldots \boxplus u_{i^*} \textbf{g}_{i^*}' \boxplus \ldots \boxplus u_k \textbf{g}_k \boxplus u_{i^*}\textbf{g}_{i^*}\boxminus u_{i^*}\textbf{g}_{i^*} ) \nonumber \\
%&\overset{\text{(b)}}{\geq}&w(u_1 \textbf{g}_1 \boxplus \ldots \boxplus u_k \textbf{g}_k) - w(u_{i^*} \textbf{g}_{i^*}' \boxminus u_{i^*} \textbf{g}_{i^*} ) \nonumber \\
%&\overset{\text{(c)}}{\geq}& (n-k+1) - \delta, \nonumber
%\end{eqnarray}
% --------- 2 column version --------
\begin{subequations}
\begin{align}
w(\textbf{v}')&=w(\textbf{uG}') \nonumber \\
&=w(u_1 \textbf{g}_1'\boxplus \!  \ldots \!  \boxplus u_{i^*} \textbf{g}_{i^*}' \boxplus \!  \ldots \! \boxplus u_k \textbf{g}_k') \nonumber \\
&\overset{\text{(a)}}{=}w(u_1 \textbf{g}_1 \boxplus \!  \ldots \!  \boxplus u_{i^*} \textbf{g}_{i^*}' \boxplus \! \ldots \! \boxplus u_k \textbf{g}_k \! \boxplus \! u_{i^*}\textbf{g}_{i^*} \! \boxminus \! u_{i^*}\textbf{g}_{i^*} ) \nonumber \\
&\overset{\text{(b)}}{\geq}w(u_1 \textbf{g}_1 \boxplus \!  \ldots \!  \boxplus u_k \textbf{g}_k) - w(u_{i^*} \textbf{g}_{i^*}' \boxminus u_{i^*} \textbf{g}_{i^*} ) \nonumber \\
&\overset{\text{(c)}}{\geq} (n-k+1) - \delta, \nonumber
\end{align}
\end{subequations}
where (a) follows from the fact that $\textbf{g}_i=\textbf{g}_i'$ $\forall i \neq {i^*}$, (b) follows from  Lemma \ref{lem:triangle2}, and
(c) follows from the Singleton lower bound. Without loss of generality, we assume that $\delta$ elements of the parity part
of $g_{i^*}$ are replaced by zeros, giving rise to $g'_{i^*}$, a codeword of $\mathcal{C}'$ of weight $n-k+1-\delta$. The theorem is proved.
\end{IEEEproof}

\section{Proof of Theorem \ref{th:gndc_mds}} \label{ap:gndc_mds}

\begin{IEEEproof}
Relating our problem with the one in Section \ref{subsec:faulty}, one fault corresponds to one inter-user channel being in outage. When that happens, the user's receiver cannot correctly decode its partner's information, so, when forming $k_2$ linear combinations to generate its $k_2$ parity-check packets, this user replaces this erroneous packet with an all-zero packet, or, equivalently, sets to zero the $k_2$ coefficients associated with this partner. This amounts to replacing by zeros the $k_2$ corresponding entries (in same row) of the parity matrix $\textbf{P}$, {\em i.e.}, $|A|=k_2$.
Since each user knows its own information, $k_2$ entries in each row of $\textbf{P}$ are immune to faults, while the other $k_2 (M-1)$ entries are subject to faults. In the worst scenario, when all the possible faults happen, the generator matrix takes the form
\begin{eqnarray} \label{eq:Gg}
\textbf{G} &=&
\left[
\begin{array}{ccc|ccc}
       &      &  & P_1    & &   \\
       & I    &  &     & \ddots  &    \\
       &      &  &     &       &  P_M
\end{array} \right],
\end{eqnarray}
where the $(k_1 \times k_2)$ submatrix $P_i$ contains the immune entries associated with user $i$. From Theorem~\ref{th:nonsingular}, we know that every submatrix of $P_i$ is nonsingular. Thus, the least minimum Hamming distance of a block code obtained from the original MDS code $\cal C$  due to the occurrence of faults is $k_2 + 1$. Nevertheless, the same minimum Hamming distance can be achieved with a much lower number of faults. From Lemma~\ref{le:0column} and Theorem~\ref{th:k2zeros}, we can see that the minimum number of faults for a code with minimum distance $k_2+1$ is $M-1$, when all of these faults occur in the same row of $\textbf{P}$, for example.

%AQUI

For all the possible minimum distances in the range $k_2+1 \leq d_{min} \leq Mk_2+1$, a sufficient condition for the worst possible scenario (the lowest number of faults that result in this minimum distance) is when all the faults are located in the same row of $\textbf{P}$, according to Lemma~\ref{le:lb}. Thus, from Lemma~\ref{le:dagg},
we can see that the larger the number of faults in a given row (and consequently the lower the minimum distance of the resulting code), the lower the composite minimum distance. This assures that the code with minimum distance $k_2+1$ is the one that generates the least composite minimum distance, which is then given by
\begin{eqnarray} \label{eq:dminGDNC}
d_{min}^{\text{comp}} &=& \min_{B_\chi \in \mathcal{B}} \left\{d_{min}(\mathcal{C}(B_\chi)) + |B_\chi|/|A| \right\} \nonumber \\
&=& (k_2 + 1) + (M - 1) \nonumber \\
&=& M + k_2
\end{eqnarray}
It is easy to see the connection between the two terms in the composed minimum distance and the exponents of $P_e$ in \eqref{eq:outagec} and \eqref{eq:newprobaa}. The proof is now complete.
\end{IEEEproof}

\section{Proof of Theorem \ref{th:quasi_mds}} \label{ap:quasi_mds}
\begin{IEEEproof}
Let $\mathcal{C}$ be an $(n,k,d_{min})$ MDS code with generator matrix  $\textbf{G}=[\textbf{I}|\textbf{P}]$. Replacing any set of entries of its parity matrix $\textbf{P}$ by zeros gives rise to an $(n,k,d_{min}')$ non-MDS code $\mathcal{C}'$ with $d_{min}'<d_{min}$, according to Theorem \ref{th:k2zeros}. Assume that $\mathcal{C}'$ is used as the non-MDS network code for the GDNC scheme. If any of the entries replaced by zero is one of the entries that are immune to fault, we can easily show that there is a pattern of $M-1$ faults that generates a codeword with weight $< k_2+1$. Thus, the composite minimum distance is $d_{min}^{\text{comp}}<M+k_2$, according to \eqref{eq:dmineff}.
\end{IEEEproof}
\section{Proof of Theorem \ref{th:q_sufficient}} \label{ap:q_sufficient}
\begin{IEEEproof}
We want to find an $(n=M(k_1 + k_2), k=Mk_1, d_{min} = Mk_2 + 1)$ MDS code over GF($q$). Let us start by choosing a field size $q \geq M(k_1 + k_2)$ as a power of a prime number. From \cite{sloane.77}, there exists an $(n' = q-1, k' = Mk_1, d'_{min} = q - Mk_1)$ RS code over GF($q$). If $q = M(k_1 + k_2) + 1$, simply use the RS code as the MDS code. If $q = M(k_1 + k_2)$, we can add a parity-check symbol and have an extended $(q, Mk_1, q - Mk_1 + 1)$ RS code over GF($q$), which can be used as the MDS code. If $q > M(k_1 + k_2) + 1$, then we can puncture the RS code in exactly $q - M(k_1 + k_2) - 1$ positions, resulting in an $(n = M(k_1 + k_2), k=Mk_1, d_{min} = Mk_2 + 1)$ code, which is MDS.
\end{IEEEproof}

\section*{Acknowledgement} This work has been supported in part by CNPq (Brazil).

% -------------------------------------------------------------------------
\IEEEtriggeratref{16}

\bibliography{IEEEabrv,biblio}
\bibliographystyle{ieeetran}

\end{document}